\documentclass{aa}

\usepackage[varg]{txfonts}
\usepackage{epsfig,graphicx,natbib,url,twoopt}
\usepackage[varg]{txfonts}
\usepackage{hyperref}          
\hypersetup{
 colorlinks=true,  
 urlcolor=blue,    
 linkcolor=red,     
 citecolor=blue 
}

\def\kms{\hbox{km$\;$s$^{-1}$}}

\def\Halpha{\mbox{H\hspace{0.1ex}$\alpha$}}
\def\Hbeta{\mbox{H\hspace{0.1ex}$\beta$}}
\def\Hgamma{\mbox{H\hspace{0.1ex}$\gamma$}}
\def\Hepsilon{\mbox{H\hspace{0.1ex}$\varepsilon$}}

\def\CaH{\ion{Ca}{ii}~H}

\defcitealias{2020A&A...641L...5J}{Paper I}
\defcitealias{2022A&A...664A..72J}{Paper II}
\begin{document}

\title{Observations of magnetic reconnection in the deep solar atmosphere in the \Hepsilon\ line}

\author {
Luc H. M. Rouppe van der Voort\inst{1,2}
\and
Jayant Joshi\inst{3}
\and
Kilian Krikova\inst{1,2}
}

\authorrunning{Rouppe van der Voort, Joshi \& Krikova}
\titlerunning{Observations of magnetic reconnection in the \Hepsilon\ line}

\institute{
  Institute of Theoretical Astrophysics,
  University of Oslo, %
  P.O. Box 1029 Blindern, N-0315 Oslo, Norway
\and
  Rosseland Centre for Solar Physics,
  University of Oslo, %
  P.O. Box 1029 Blindern, N-0315 Oslo, Norway
\and
  Indian Institute of Astrophysics, 
  II Block, Koramangala, Bengaluru 560 034, India
}

\date{submitted to A\&A Dec 15, 2023 / accepted Jan 21, 2024}


\abstract
{
Magnetic reconnection in the deep solar atmosphere can give rise to enhanced emission in the Balmer hydrogen lines, a phenomenon known as Ellerman bombs (EBs). It is most common to observe EBs in the \Halpha\ and \Hbeta\ spectral lines. High quality shorter wavelength Balmer line observations of EBs are rare but have the potential to provide the most highly resolved view on magnetic reconnection. }
{
We aim to evaluate the \Hepsilon\ 3970~\AA\ line as an EB diagnostic by analyzing high quality observations in different Balmer lines.
}
{
Observations of different targets and viewing angles were acquired with the Swedish 1-m Solar Telescope. These observations sample EBs in different environments: active regions, quiet Sun, and the penumbra and moat of a sunspot. 
We employed an automated detection method for quiet Sun EBs based on $k$-means clustering.
}
{
Ellerman bombs in the \Hepsilon\ line show similar characteristics as in the longer wavelength Balmer lines: enhanced intensity as compared to the surroundings, rapid variability, and flame-like morphology. 
In a 24 min quiet Sun time series, we detected 1674 EBs in the \Hepsilon\ line which is 1.7 times more than in \Hbeta.
The quiet Sun EBs measured in \Hepsilon\ are very similar as in \Hbeta: they have similar lifetimes, area, brightness, and spatial distribution. Most of the EBs detected in \Hepsilon\ are closer to the limb than their \Hbeta\ counterparts. This can be explained by the \Hepsilon\ line core EB emission being formed higher in the atmosphere than the \Hbeta\ EB wing emission. 
} 
{
We conclude that the \Hepsilon\ line is well suited for studying EBs and consequently measure the dynamics of magnetic reconnection in the solar atmosphere at the smallest scales. Our findings suggests that the deep atmosphere in the quiet Sun may host more than 750,000 reconnection events with EB signature at any time. That is significantly more than what was found in earlier \Hbeta\ observations.
}

\maketitle

\section{Introduction}
\label{sec:introduction}

The hydrogen Balmer lines can be used as a tracer of magnetic reconnection in the deep solar atmosphere. 
Enhanced emission in these spectral lines can be seen in sites where opposite magnetic polarities are in close proximity, flux cancellation is observed, and magnetic reconnection is thought to take place. 
The occurrence of concentrated sites with strong emission in the \Halpha, \Hbeta, and \Hgamma\ lines in active regions were first reported by 
\citet{1917ApJ....46..298E} 
and the phenomenon is now commonly known as Ellerman bombs (EBs). 
The use of EB emission as a proxy for magnetic reconnection is supported by a vast body of observations in the current age of high resolution solar observations
\citep[see, e.g.,][]{2002ApJ...575..506G, 
2004ApJ...614.1099P, 
2006ApJ...643.1325F, 
2007A&A...473..279P, 
2008PASJ...60..577M, 
2008ApJ...684..736W, 
2010ApJ...724.1083G, 
2016ApJ...823..110R}. 

The characteristic EB spectral profile has enhanced wings that appear in emission (with peak emission in both wings around 40~\kms\ Doppler offset) and line core absorption that has similar low intensity level as the surroundings, this profile is sometimes described as moustache like
\citep{1964ARA&A...2..363S}. 
The unperturbed line core is a sign that the site that is the origin of the enhanced emission is situated below the chromospheric canopy of fibrils. This is confirmed in high-resolution imaging spectroscopy
\citep{2011ApJ...736...71W, 
2013ApJ...774...32V, 
2013JPhCS.440a2007R, 
2013ApJ...779..125N}. 
Under inclined viewing angle, observing regions away from disk center, \Halpha\ wing images show EBs as 1--2~Mm, bright, upright flames that flicker rapidly on a time scale of seconds and are rooted in the deep photosphere
\citep{2011ApJ...736...71W, 
2013JPhCS.440a2007R, 
2015ApJ...798...19N}. 
The rapid variability and flame substructure have been interpreted as a sign of fast reconnection and the formation of plasmoids \citep{2023A&A...673A..11R, 
2017ApJ...851L...6R}. 
Numerical simulations have reproduced the formation of typical EB spectral profiles from flame-like structures that are regions with reconnection along up-right, thin current sheets that are rooted in intergranular lanes and extend through the low solar atmosphere
\citep{2017ApJ...839...22H, 
2017A&A...601A.122D, 
2019A&A...626A..33H}. 

Ellerman bombs are most commonly observed in active regions with fast flow patterns that move magnetic fields such as in emerging flux regions. 
High spatial resolution 
observations of tiny EB-like flames in quiet Sun demonstrated that the EB phenomenon is not exclusive for active regions 
\citep{2016A&A...592A.100R, 
2017ApJ...845...16N, 
2018MNRAS.479.3274S, 
2023ApJ...944..171B}. 
Recently, \citet[][hereafter \citetalias{2020A&A...641L...5J}]{2020A&A...641L...5J} 
and
\citet[][\citetalias{2022A&A...664A..72J}]{2022A&A...664A..72J} 
found that quiet Sun EBs (QSEBs) are much more ubiquitous in new, high quality \Hbeta\ observations.
The shorter wavelength \Hbeta\ line allows for higher spatial resolution and contrast 
than \Halpha\ 
and facilitates detection of smaller and weaker EB events. 
%
They found that half of the QSEBs in \Hbeta\ have a width that is smaller than 0\farcs19 which is a spatial scale that is very challenging to resolve in \Halpha.
%
Their analysis suggested that about half a million QSEBs are present in the solar atmosphere at any time. 
High spatial resolution \Hbeta\ observations of sunspots showed that penumbrae are filled with large numbers of EB flames
\citep{2021A&A...648A..54R}. 
The high density of penumbral EBs (PEBs) suggests that magnetic reconnection is ubiquitous in the deep atmosphere of sunspot penumbrae. 

The remarkable enhanced EB visibility in the recent \Hbeta\ observations raise the question whether shorter wavelength Balmer lines show even finer detail and higher EB occurrence in the solar atmosphere. 
The \Hepsilon\ line at 3970~\AA\ is only about 1.5~\AA\ from the \CaH\ line core.
Its vicinity to the strong \CaH\ line has some advantages from an instrumentation point of view but also poses some challenges for interpretation.
Recently, \citet{2023A&A...677A..52K} 
presented a detailed study of the spectral line formation of \Hepsilon. 
The \Hepsilon\ line is most often a weak absorption feature against the extended \CaH\ wings and 
shows the reversed granulation intensity pattern that is formed a few hundred km above the photosphere.
This means that the \Hepsilon\ line core reflects mostly the \CaH\ background radiation. 
However, \citet{2023A&A...677A..52K} 
also presented observations of small regions in quiet Sun with \Hepsilon\ in emission indicating heating.
Here, we present a systematic overview of high quality observations of different targets to evaluate the \Hepsilon\ line as an EB diagnostic. 

\begin{figure*}[!ht]
\centering
\includegraphics[width=\textwidth]{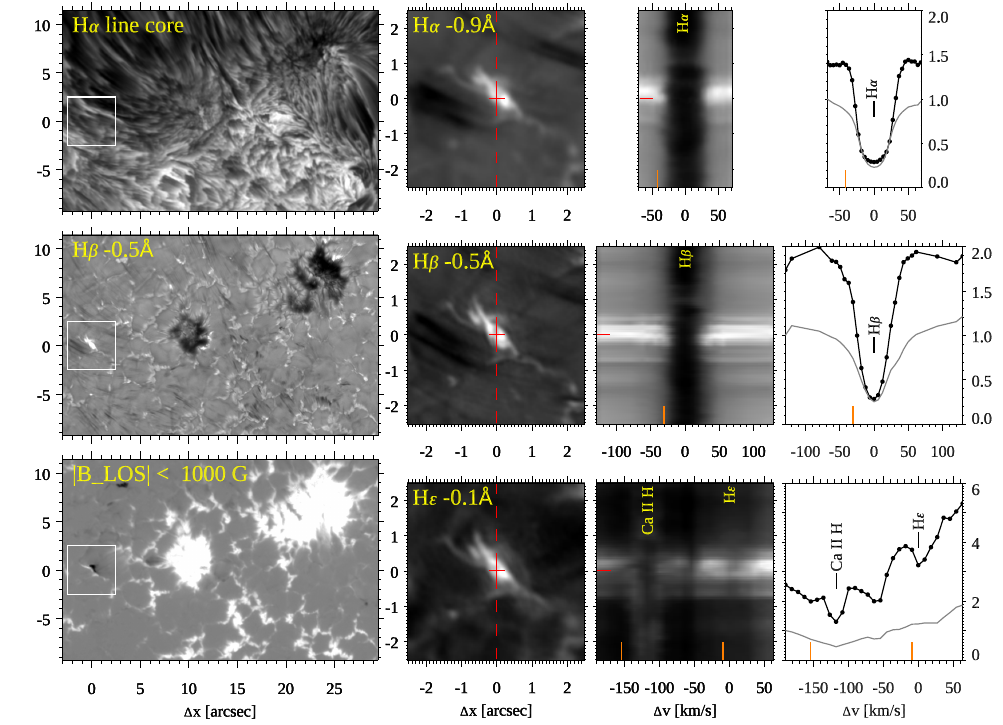} \\
\includegraphics[width=\textwidth]{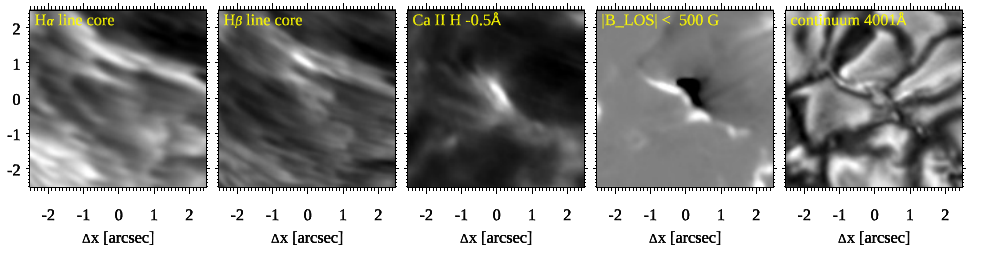} \\
\caption{\label{fig:textbook}%
Strong EB in an active region observed on 11-Aug-2020. The larger rectangular panels at the left of the top three rows show the larger context area in \Halpha\ line core, \Hbeta\ blue wing, and a magnetogram. The white square is centred on the EB and marks the area shown at larger magnification in the other square images. The red dashed line in the Balmer blue wing images (\Halpha, \Hbeta, \Hepsilon) marks the artificial slit for which the corresponding spectrogram ($\lambda y$-diagram) is shown to the right. The EB spectral profiles, marked by red horizontal dashes, are shown in the right panels as solid black lines with the observed sampling points marked as small filled circles. The thin gray profiles are reference spectral profiles averaged over the full region shown in the left context images. Vertical orange dashes mark the wavelength positions of the Balmer wing images, and the \CaH\ blue wing image in the bottom row. 
Two animations are available in the online material: one showing the temporal evolution of the middle rows of this figure
(see \url{http://tsih3.uio.no/lapalma/subl/heps/rouppe_heps_fig01_timeevol.mp4}), and one showing spectral line scans of the top three rows
(see \url{http://tsih3.uio.no/lapalma/subl/heps/rouppe_heps_fig01_linescan.mp4}). 
}
\end{figure*}

\begin{figure*}[!ht]
\centering
\includegraphics[width=\textwidth]{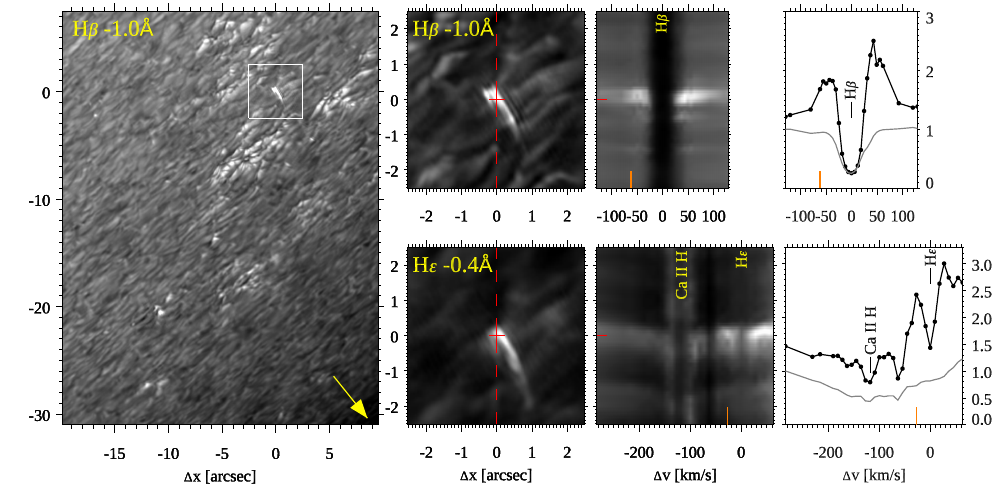} \\
\includegraphics[width=\textwidth]{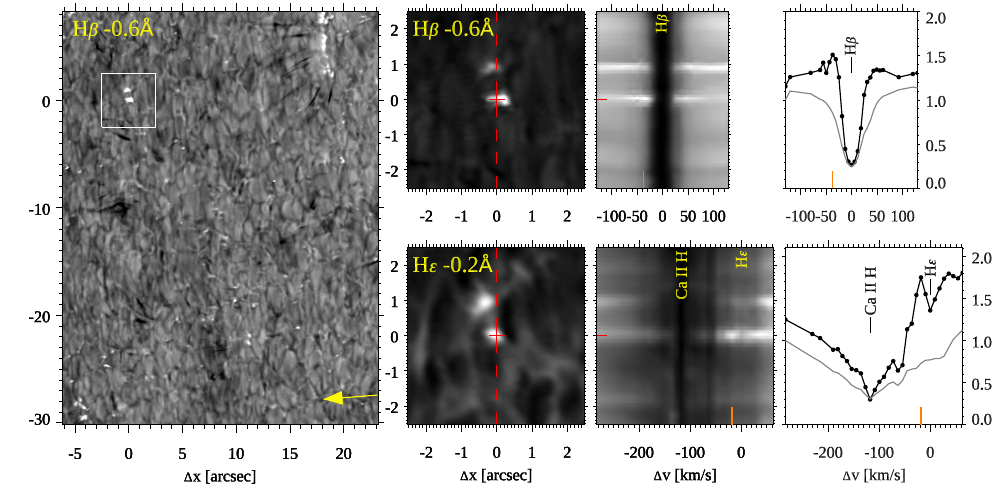} \\
\caption{\label{fig:examples}%
Examples of EBs in a small active region (top, 13-Aug-2020) and in quiet Sun (bottom, 15-Aug-2020). Both examples are close to the limb (top: $\mu=0.32$, bottom: $\mu=0.49$). The small white square in the overview images at left are centered on the EBs shown at larger magnification in the \Hbeta\ and \Hepsilon\ wing images in the next column. A yellow arrow in the lower-right corner of the overview image shows the direction towards the closest limb. The $\lambda y$ spectrogram is shown for the vertical red dashed line in the center of the wing images and crosses the EB. The spectral profiles are shown for the center position that is marked with the short horizontal red dash in the spectrogram. The thin gray profiles are reference spectral profiles averaged over the full region shown in the left context images. 
Animations that show the full spectral line scans of the two examples are available in the online material
(see \url{http://tsih3.uio.no/lapalma/subl/heps/rouppe_heps_fig02_ARlimb_linescan.mp4} and \url{http://tsih3.uio.no/lapalma/subl/heps/rouppe_heps_fig02_QSEB_linescan.mp4}). 
}
\end{figure*}

\begin{figure*}[!ht]
\centering
\includegraphics[width=\textwidth]{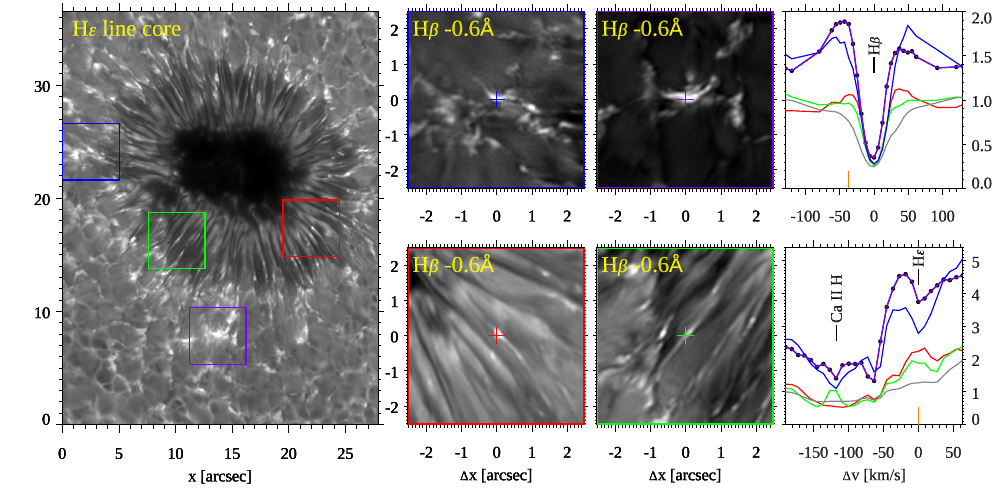} \\
\caption{\label{fig:sunspot}%
Ellerman Bombs in the sunspot penumbra and around the sunspot in the moat observed on 07-Aug-2020. Colored squares in the overview \Hepsilon\ line core image at left are centered on EB examples that are shown at higher magnification in \Hbeta\ wing in the center. The blue and purple line profiles in the spectral plots at right are for the two EBs outside the sunspot. The red and green profiles are PEBs in the penumbra. 
}
\end{figure*}

\begin{figure*}[!ht]
\centering
\includegraphics[width=\textwidth]{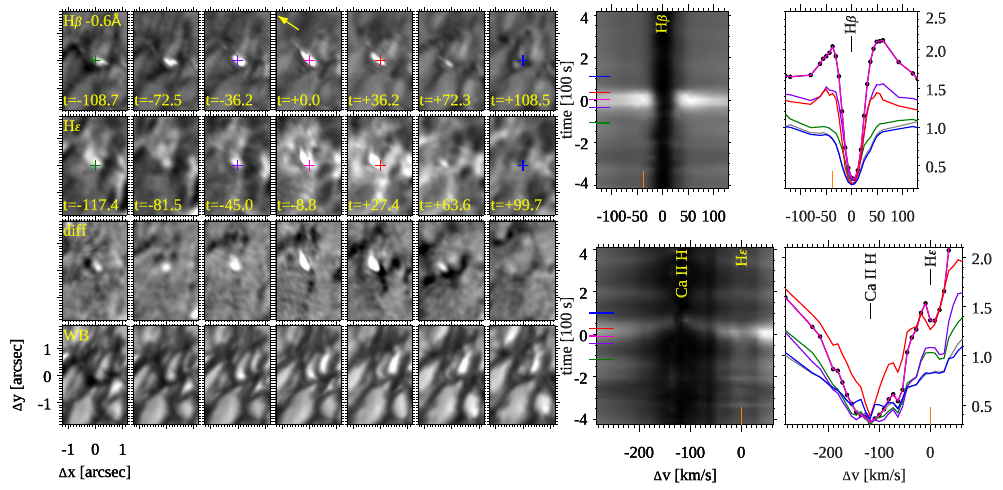} \\
\caption{\label{fig:timeseries}%
Temporal evolution of a QSEB observed at $\mu=0.54$ on 16-Aug-2020. On the left, a series of small images are centered on a QSEB in, from top to bottom row: \Hbeta\ wing, \Hepsilon\ line center, difference \Hepsilon\ $-$ \CaH\ blue wing, and WB 4846~\AA. The $\lambda t$-diagrams to the right show the spectral evolution at the pixel location marked with a cross in the \Hbeta\ wing and \Hepsilon\ images. Colored markers indicate the times for which spectral profiles are shown at right. 
An animation of the this figure is available in the online material (see \url{http://tsih3.uio.no/lapalma/subl/heps/rouppe_heps_fig04_QSEB.mp4}).
}
\end{figure*}



\section{Observations}
\label{sec:observations}

\begin{table*}[h]
\caption{Overview over the data sets analysed in this study.}              
\label{table:1}      
\centering                                      
\begin{tabular}{c c c c c c c}          
\hline\hline                        
Date & Time (UTC) & Target & $(X,Y)$ [\arcsec] & $\mu$ &  $r_0$ [cm] & Fig. \\    
\hline                                   
2020-08-07 & 13:02:09 & AR12770 & ($-$412, 285) & 0.85 & 29 (21) & \ref{fig:sunspot} \\
2020-08-11 & 09:28:04 & AR12770 & (318,295)     & 0.89 & 23 (10) & \ref{fig:textbook} \\
           & 09:23:24 -- 09:39:02 & & & & 33 (11) & \ref{fig:textbook} (movie)\\  
2020-08-13 & 07:58:51 & AR      & ($-$750, $-$502) & 0.32 & 26 (9) & \ref{fig:examples} \\
2020-08-15 & 09:00:26 & CH      & (0, 830)      & 0.49 & 44 (12) & \ref{fig:examples} \\
2020-08-16 & 08:32:20 -- 08:36:10 & QS & (515, 619) & 0.54 & 45 (12) & \ref{fig:timeseries} \\
           & 08:17:01 -- 08:41:12 &    &            &      & 51 (13) & \ref{fig:stats}--\ref{fig:seeing} \\
\hline                                             
\end{tabular}
\tablefoot{%
Target types: active region (AR), coronal hole (CH), or quiet Sun (QS). 
$(X,Y)$: approximate pointing coordinates. 
$\mu = \cos \theta$ with $\theta$ the observing angle.
$r_0$: maximum value of the Fried's parameter for the ground-layer seeing, the value between parentheses covers both ground-layer and high-altitude seeing. 
The last column shows for which figures the different data sets are used.
}
\end{table*}

\begin{figure*}[!ht]
\centering
\includegraphics[width=0.9\textwidth]{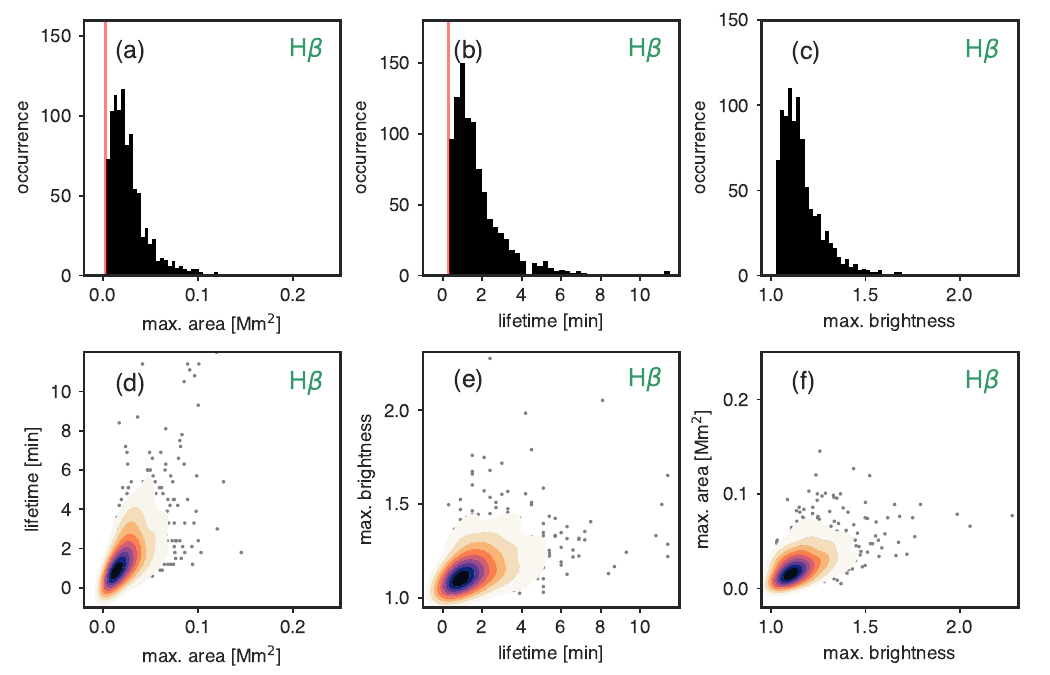} \\
\includegraphics[width=0.9\textwidth]{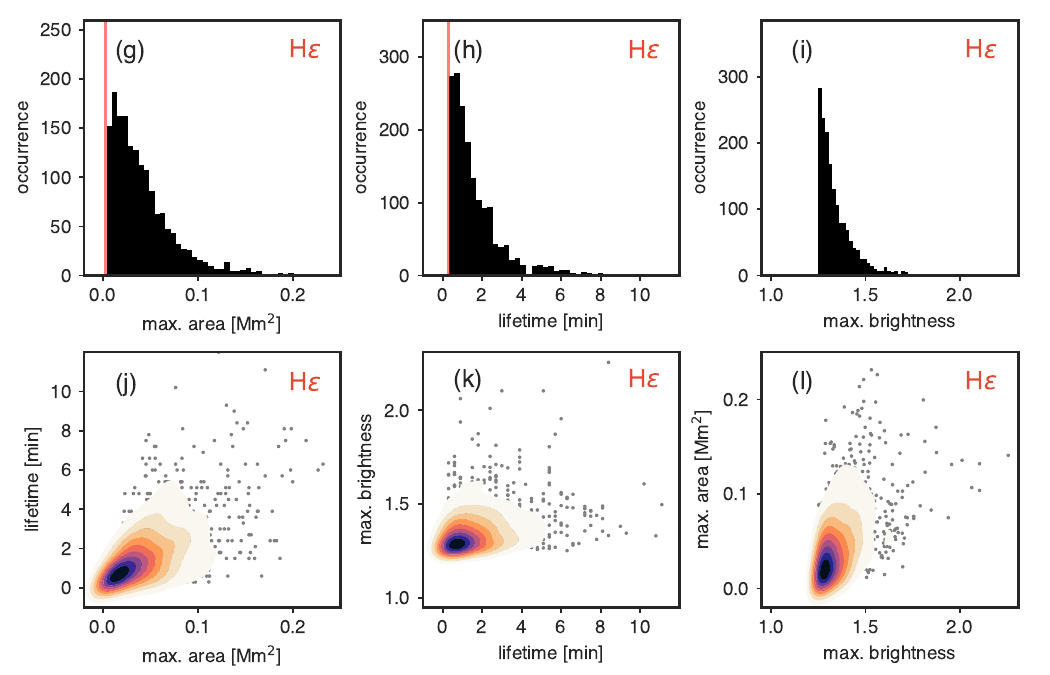} \\
\caption{\label{fig:stats}%
Statistical properties of QSEBs detected in the 16-Aug-2020 24~min time series. The total number of \Hbeta\ QSEBs is 961 and of \Hepsilon\ QSEBs 1674. The filled black histograms in panels (a)--(c) and (g)--(i) represent the maximum area, lifetime, and maximum brightness distributions, respectively. The vertical red line marks the lower limit set by sampling: 0.0008 Mm$^2$ (one pixel) in area (panels (a) and (g)) and 18~s in lifetime (panels (b) and (h)). In panels (d)--(f) and (j)--(l), multivariate JPDFs and scatter plots between the maximum area, lifetime, and maximum brightness are shown. The dark blue shade of the JPDFs indicates the highest density occurrence, and the lighter orange shaded regions represent the low-density distribution. 
}
\end{figure*}

\begin{figure}[!ht]
\centering
\includegraphics[width=0.45\textwidth]{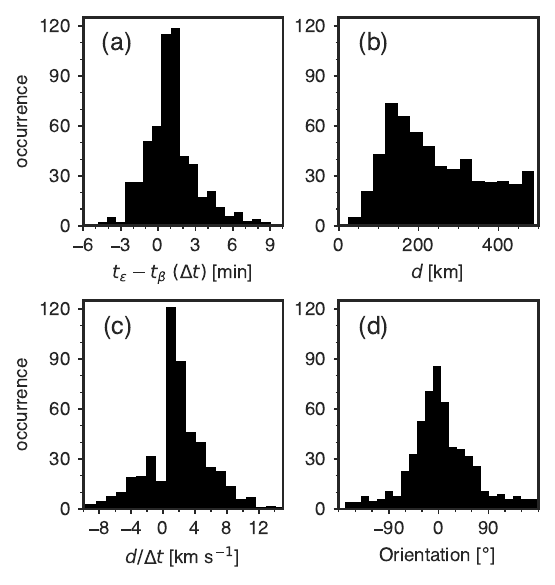} \\
\caption{\label{fig:beta_vs_eps}%
Time difference, distance, average propagation speed, and orientation of the brightening of QSEBs in H$\epsilon$ with respect to their counterparts in H$\beta$. Measurements are based on when the QSEB appeared first in the respective spectral line. Positive values of the propagation speed $d/\Delta t$ implies that the QSEB occurred in \Hbeta\ first. The orientation angle is measured against the direction toward the closest limb. An orientation of 0\degr\ means that the QSEB in \Hepsilon\ is closer to the limb than the QSEB measured in \Hbeta. The total number of QSEBs measured in both \Hepsilon\ and \Hbeta\ is 561. 
}
\end{figure}

\begin{figure}[!ht]
\centering
\includegraphics[width=0.48\textwidth]{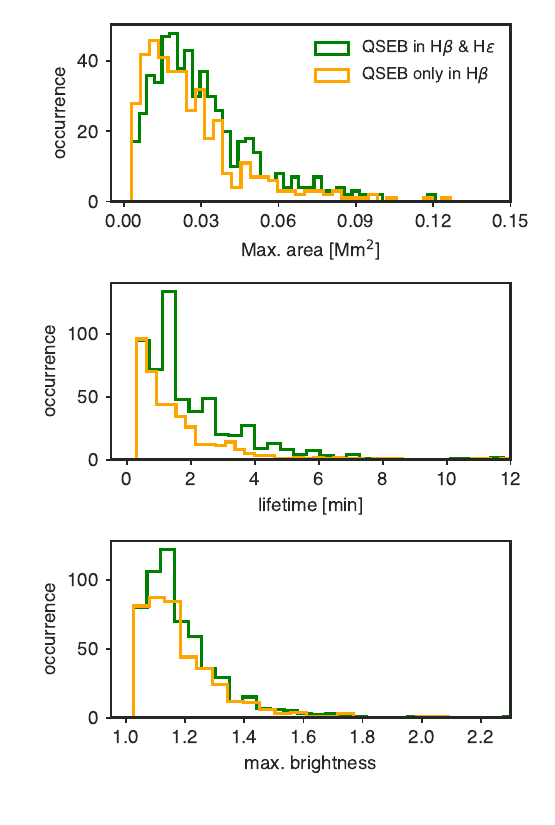} \\
\caption{\label{fig:stat_comp}%
Comparison of statistical properties measured in \Hbeta\ between QSEB events detected both in the \Hbeta\ and \Hepsilon\ lines and those which are found only in the \Hbeta\ line. The total number of QSEBs detected in both \Hepsilon\ and \Hbeta\ is 561. The total number of QSEBs detected only in \Hbeta\ is 400.
}
\end{figure}

\begin{figure*}[!ht]
\centering
\includegraphics[width=0.48\textwidth]{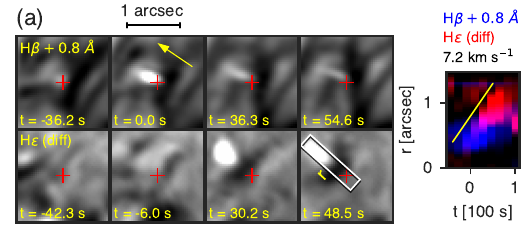}
\includegraphics[width=0.48\textwidth]{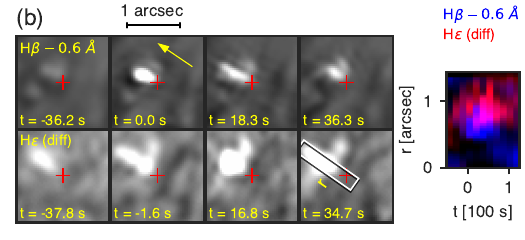}\\
\includegraphics[width=0.48\textwidth]{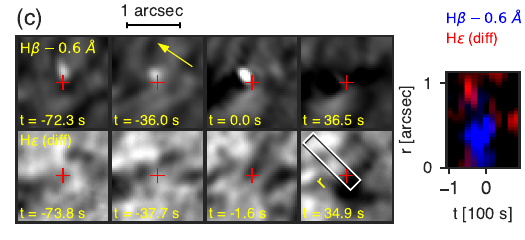}
\includegraphics[width=0.48\textwidth]{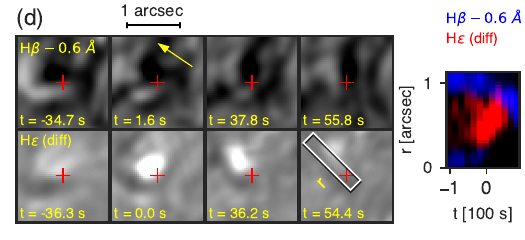}\\
\caption{\label{fig:roi_offset}%
Examples of the temporal evolution of QSEBs. 
(a) QSEB observed first in \Hbeta\ and then in \Hepsilon. (b) QSEB observed first in \Hepsilon\ and then in \Hbeta. (c) QSEB observed only in \Hbeta. (d) QSEB observed only in \Hepsilon. The bottom row of images show difference (\Hepsilon\ $-$ \CaH\ blue wing) images where white shows emission in \Hepsilon. Each image is scaled individually. The bottom right panels show a rectangular box along which a space-time diagram along distance $r$ is shown in the right panels. The yellow arrows indicate the direction towards the limb so that $r = 1$ is closer to the limb. 
}
\end{figure*}

\begin{figure*}[!ht]
\centering
\includegraphics[width=0.95\textwidth]{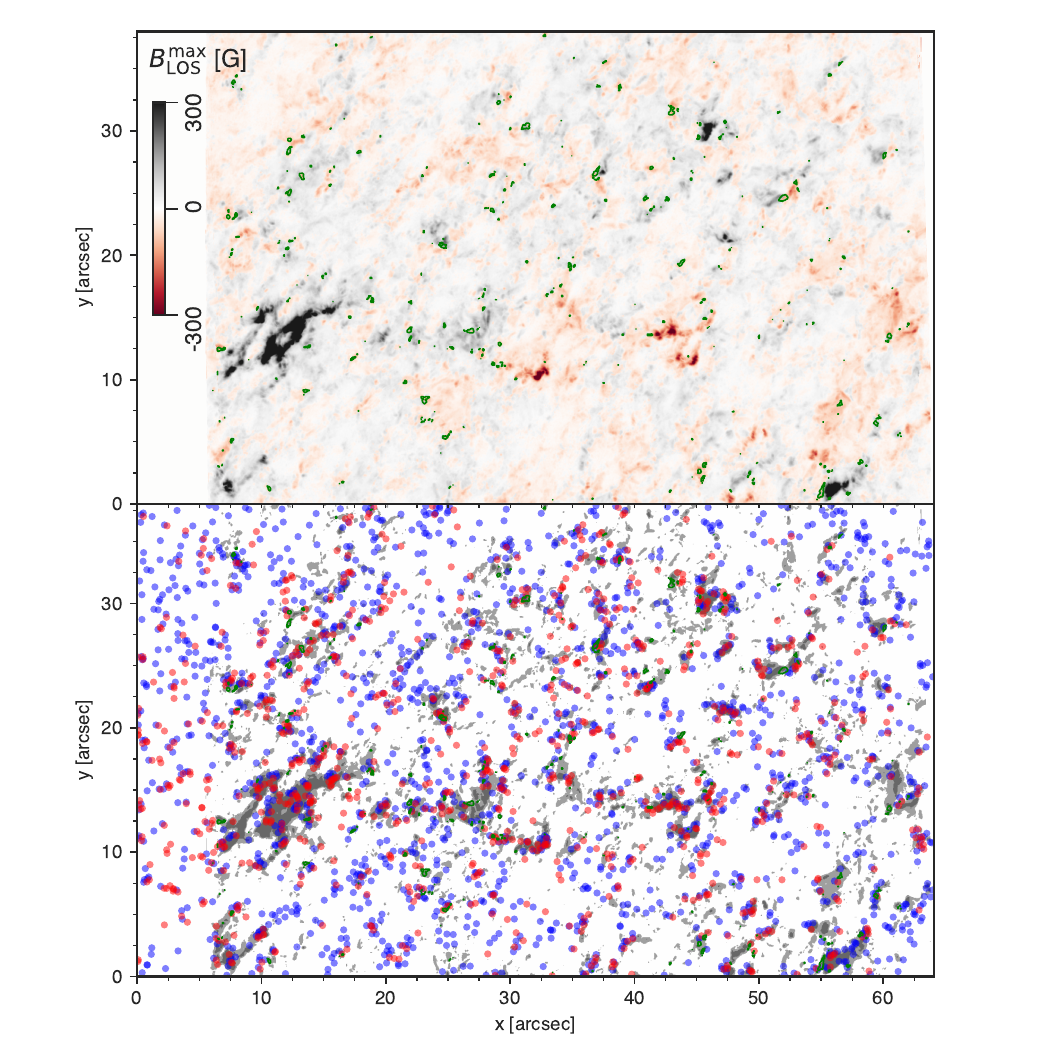} \\
\caption{\label{fig:spatial_dist}%
Spatial distribution of QSEBs and their magnetic environment. The top panel shows, at each pixel, the extremum of $B_\mathrm{LOS}$ over the full 24~min duration of the time series. Green contours mark pixels that have $|B_\mathrm{LOS}| > 50$~G for both polarities during the time series. The bottom panel shows QSEB detections in \Hbeta\ (red) and \Hepsilon\ (blue). There are 961 QSEBs in \Hbeta\ and 1674 QSEBs in \Hepsilon. The shaded background marks regions where $|B_\mathrm{LOS}^\mathrm{max}| > 100$~G (dark gray) and $50 < |B_\mathrm{LOS}^\mathrm{max}| < 100$~G (light gray). The CRISP FOV is slightly narrower than for CHROMIS so that the region for $x<6\arcsec$ is not covered by the magnetic field map.
}
\end{figure*}

%
The observations were obtained with the CHROMIS 
and CRISP
\citep{2008ApJ...689L..69S} 
imaging spectro(polari)meters at the Swedish 1-m Solar Telescope 
\citep[SST, ][]{2003SPIE.4853..341S} 
in August 2020.
With CHROMIS, we cycled through a multi-line program sampling the \Hbeta, \CaH, and \Hepsilon\ spectral lines. 
The \Hbeta\ line was sampled at 27 line positions, between $\pm2.1$~\AA, with 0.10~\AA\ steps between $\pm1.0$~\AA, and coarser in the outer wings, avoiding strong line blends. The CHROMIS transmission bandwidth at \Hbeta\ is 0.10~\AA.
The prefilter that was used for \CaH\ has center wavelength 3968.8~\AA\ and bandwidth 4.1~\AA\ and allows for sampling of the \Hepsilon\ line at 3970.1~\AA\ (at about +1.5~\AA\ offset from \CaH\ line center). 
We sampled the \CaH\ line at 29 line positions from $-0.96$ to $+2.40$~\AA\ with steps of 0.12~\AA. 
The CHROMIS transmission bandwidth at \CaH\ is 0.12~\AA. 
At $+2.40$~\AA, the prefilter transmission is about 50\% of peak transmission. 
With this program we cover extreme \Hepsilon\ emission profiles out to about +0.9~\AA\ from nominal \Hepsilon\ line center. 
Later in the observing campaign, from 13 Aug 2020, we added 3 line positions in the \CaH\ blue wing (at $-2.20, -1.50$, and $-1.30$~\AA). 
These extra blue wing positions allow for simple subtraction images with the \Hepsilon\ position in the \CaH\ red wing which effectively show sites with enhanced emission or absorption in the \Hepsilon\ line. 
In addition to sampling the \CaH\ line, a continuum position at 4001~\AA\ was observed. This required a different CHROMIS prefilter but was observed with the same wideband (WB) filter as for \CaH\ (center wavelength 3950~\AA\ and bandwidth 13.2~\AA). 
For \Hbeta\ the WB filter had center wavelength 4846~\AA\ and bandwidth 6.5~\AA. 
Both WB channels effectively sample the photosphere.
The temporal cadence of the CHROMIS observations was about 17~s (18~s from 13 Aug 2020). 
CHROMIS has a pixel scale of 0\farcs038, and a FOV of 66\arcsec $\times$ 42\arcsec.

With CRISP, we sampled the \Halpha, \ion{Fe}{i}~6173~\AA{}, and \ion{Ca}{ii}~8542~\AA{} spectral lines at a cadence of 40~s.
CRISP sampled the \Halpha{} line at 31 line positions between $\pm$1.5~\AA{} with 0.1~\AA\ steps. 
The \ion{Fe}{i}~6173~\AA{} line was observed with polarimetry and was sampled at 13 line positions (between $\pm0.16$~\AA\ with 0.04~\AA\ steps, and further at $\pm0.24$~\AA\ and $\pm0.32$~\AA) plus the continuum at $+0.68$~\AA\ from the nominal line core. 
For each line position and polarisation state, 8 exposures were acquired that were used for image restoration (i.e., a total of 448 exposures per spectral line scan). 
The noise level in the restored Stokes V/I$_\textrm{cont}$ maps was estimated to be 2$\times$10$^{-3}$.
Furthermore, spectropolarimetric observations were acquired in the \ion{Ca}{ii}~8542~\AA\ line at 20 line positions between $-1.68$ and $+2.38$~\AA. 
Maps of the magnetic field strength along the line of sight ($B_\textrm{LOS}$) were derived from Milne-Eddington inversions of the \ion{Fe}{i}~6173~\AA\ observations using the inversion code developed by 
\citet{2019A&A...631A.153D}. 
We estimate the noise level in the $B_\mathrm{LOS}$ maps to be 6~G. This was measured as the standard deviation in a very quiet region in the quiet Sun time series of 16-Aug-2020. 

The data was processed from raw exposures to science-ready data cubes using the SSTRED reduction pipeline \citep{2015A&A...573A..40D, 
2021A&A...653A..68L}. 
An important step in the processing pipeline is the application of image restoration with the multi-object multi-frame blind deconvolution 
\citep[MOMFBD, ][]{2005SoPh..228..191V} 
method. 
Each of the CRISP spectral lines and CHROMIS \Hbeta\ data were processed separately while \Hepsilon\ and 4000~\AA\ continuum was processed together with \CaH\ (the data was effectively separated by filter in the WB channel).
One of the final steps in the pipeline was alignment between the spectral line data cubes. This was done by cross-correlation of the WB channels that show similar photospheric scenes.
The CHROMIS FOV and temporal cadence served as reference to which the lower resolution CRISP data (pixel scale 0\farcs058) was matched in space (CRISP FOV about 59\arcsec $\times$ 59\arcsec) by linear interpolation and in time by nearest-neighbor sampling. 
The alignment of the data included destretching to account for residual seeing-induced image deformation that was not accounted for by image restoration. 

Details of the different data sets are provided in Table~\ref{table:1}. 
This includes measurements of the seeing quality in terms of the Fried's parameter $r_0$ as provided by the SST adaptive optics wavefront sensor
\citep[see][]{2019A&A...626A..55S}. 
High spatial resolution is required to resolve the smallest EBs and fine structure in the larger EBs. 
We therefore selected spectral scans and time sequences during the best seeing conditions from the multi-day observing campaign. 

For a detailed comparison between QSEBs in \Hbeta\ and \Hepsilon, we selected a 24~min time series from the 16-Aug-2020 observation at $\mu=0.54$ (81 time steps).
The seeing was very good and stable during the full duration of the series with only one time step with $r_0 = 8$~cm while 92\%\ of the time steps $r_0 > 15$~cm (ground-layer seeing).  
For the full atmosphere seeing, $r_0 > 8$~cm for 93\%\ of the time.

\section{Methods}
\label{sec:method}

To gather statistics from the 24~min time series, we used an automated detection method to identify and track QSEBs. 
To identify spectral signatures of QSEBs in the \Hbeta\ and \Hepsilon\ spectral data, we used the $k$-means clustering algorithm \citep{everitt_1972} 
In particular, we used the 
$k$-means++ \citep[][]{arthur2007k} 
implementation in scikit-learn which employs an optimized method for initialization. 
The basics of the methods are discussed in detail in 
\citetalias{2022A&A...664A..72J} 
and we concentrate here on some of the differences in the methods we employ here. 

Before performing the $k$-means clustering the \Hbeta{} profiles were normalized by the average of the far-wing intensities.  
Then we applied principle component analysis (PCA) to reduce the dimensionality of the \Hbeta\ data set. 
The first ten PCA components explain 90\% of the total variability in the data set, and these first ten components were further used for the $k$-means clustering.  
We clustered the PCA manipulated \Hbeta{} data into 100 groups ($k=100$). 
While the $k$-means clustering was performed on the PCA manipulated data, representative profiles (RPs) corresponding to each cluster were calculated from the original \Hbeta\ profiles.
Out of 100 RPs we found 15 RPs with QSEB like spectral signatures and those RPs are shown in Fig.~\ref{fig:RPs_hbeta} in the Appendix.
For a detailed discussion of using PCA preprocessing before $k$-means clustering, see, e.g., \citet{JonasMaster2022}.

While the fully observed spectral range was used for clustering the \Hbeta\ profiles, for $k$-means clustering in \Hepsilon\ we concentrated on the 13 spectral positions around \Hepsilon\ line center, from $-0.54$ to $+0.90$~\AA. 
Prior to $k$-means clustering, we normalized by 
the intensity level of the \CaH\ blue wing position at $-1.5$~\AA. 
From a total of 100 clusters ($k=100$), we identified 25 clusters to have \Hepsilon\ QSEB profiles, these are shown in Fig.~\ref{fig:RPs_heps}.

We selected ten time steps with good seeing conditions out of the total of 81 time steps to train the \Hbeta\ and \Hepsilon\ $k$-means models. 
Apart from these selected time steps, we incorporated pixels exhibiting potential QSEB signatures from the entire time series. 
The inclusion of these pixels was implemented by intensity thresholds, and during the training of $k$-means models, they were assigned a weight of four times higher than the pixels from the ten best scans. 
The derived models were used to predict the closest RP for each pixel in the complete time series.

For the detection of QSEB events in the different time steps and tracking them over time, we closely follow the methods described in 
\citetalias{2022A&A...664A..72J}. 
This includes the three-dimensional (3D) morphological closing operation method to connect pixel areas with QSEB RPs and 3D connected component labeling \citep{labeling_1996} 
to uniquely label events that are connected in space and time. 
We excluded the events that have a lifetime shorter than two time steps (36~s) and have maximum area less than five pixels. 
This means that both single-time step large events and small events living $\ge$2 time steps are considered as genuine QSEBs. 
Some QSEB detections in one spectral line were close in space and time to a detection in the other line and can be regarded as a single event detected in both lines. 
We consider an QSEB event as connected across
the two spectral lines if the spatial offset and the temporal gap between their respective counterparts are smaller than 500~km and 162~s. See the Appendix for a more detailed discussion of connecting events. 

\section{Results}
\label{sec:results}

\subsection{Ellerman bomb characteristics in \Hepsilon}

Figure~\ref{fig:textbook} presents a strong EB in an active region. 
It is a fine example of an EB observation that displays many of the typical EB characteristics: strong \Halpha\ wing emission, visible in the wing as a compact brightening of about 1\arcsec\ size but covered by chromospheric fibrils in the line core, occurring at the interface between opposite polarity magnetic field patches, and invisible in the 4001~\AA\ continuum channel. 
The accompanying movie almost shows the full $\sim$15~min lifetime of the event during which the EB displays rapid variability.
The intensity signal in \Hbeta\ wing is strongly enhanced and varying from frame to frame and at the same time the morphology is rapidly changing. Some thin linear extensions appear to emerge from the EB site with sometimes detaching blob-like brightenings. This morphology is compatible with the flickering flame-like behaviour that can be observed for EBs at more slanted view further towards the limb (this AR was observed with $\mu=0.89$). 

This event is associated with magnetic flux emergence where the emerging negative polarity is crashing into pre-existing positive polarity concentrations. 
The emergence of magnetic flux was accompanied with occurrence of elongated, so-called anomalous granulation. This can be best viewed in the movie associated with Fig.~4 in 
\citet{2023A&A...673A..11R} 
that was made from this observation. 

Both the \Halpha\ and \Hbeta\ spectral profiles in Fig.~\ref{fig:textbook} show the classic EB profile with strong enhanced wings. 
The \Hepsilon\ line is also strongly enhanced. It appears that the \Hepsilon\ line center itself  is in absorption against enhanced \Hepsilon\ wings. It is however difficult to distinguish the \Hepsilon\ line profile from the \CaH\ profile. The \CaH\ line itself is enhanced too, with enhanced H$_3$ line center and enhanced H$_2$ peaks as compared to the reference profile. The limited coverage of the \CaH\ blue wing for this observation makes it difficult to see to what extent the \Hepsilon\ line is enhanced against the far \CaH\ wings. 
In any case, the morphology of the EB in \Hepsilon\ blue wing is very similar as compared to the corresponding \Halpha\ and \Hbeta\ blue wing images. This is also clear from the associated movie that shows that the EB evolution in \Hepsilon\ blue wing is almost identical as in \Hbeta\ blue wing. 
The morphology of the EB in \CaH\ blue wing is slightly different but there is a clear bright linear feature at the EB site. 

Figure~\ref{fig:examples} shows two examples of EBs in magnetically less active areas. Both are further towards the limb as compared to the EB in Fig.~\ref{fig:textbook} and the apparent upright flame morphology is clear. 
In both examples, the flames appear to be aligned in the limb direction.  
The top example shows a tall EB flame in a small active region that has an apparent length of more than 1\farcs7 in \Hbeta\ wing. This corresponds to a geometric length of more than 1300~km assuming a strictly vertical structure and taking the slanted viewing angle into account.  
The bottom example is a QSEB with an apparent length of about 0\farcs7 (corresponding to an almost 600~km tall vertical structure). 
Both EBs are clearly visible in both the \Hbeta\ and \Hepsilon\ wing images and show similar morphology in both lines. 
In the bottom example, there is another QSEB about 1\arcsec\ above the QSEB in the center. It appears to be largest in the \Hepsilon\ wing image. 
For these observations we sampled more of the \CaH\ blue wing which makes it more clear to see that the \Hepsilon\ line has enhanced intensity as compared to the \CaH\ wing. 
Both examples have pronounced central absorption at \Hepsilon\ line center. 

From close inspection of the zoomed wing images in Fig.~\ref{fig:examples}, it can be seen that there is a subtle spatial offset of the EBs between the two spectral lines: the EB in \Hepsilon\ is slightly offset in the direction towards the limb. 
For the tall flame in the active region in the top example, the red cross is centered on the base of the EB in \Hepsilon\ while the corresponding EB base in \Hbeta\ wing is a few 0\farcs1's towards the top left. 
The same can be seen for the two QSEBs in the bottom example: the QSEB brightenings in \Hepsilon\ are slightly shifted to the left, in the direction towards the limb. 
The spatial offset for QSEBs in the two lines is analysed in further detail below. 

Figure~\ref{fig:sunspot} shows examples of EBs in and around a sunspot. The EBs in the sunspot moat outside the penumbra are very bright in the \Hbeta\ wing images and have strongly enhanced intensity in the both the \Hbeta\ and \Hepsilon\ wings. 
For both EBs, the \Hepsilon\ line clearly has enhanced emission compared to the \CaH\ blue wing and an absorption feature at \Hepsilon\ line center.
The EBs in the penumbra, or PEBs, have the characteristic EB spectral line shape in \Hbeta\ but with much weaker wing enhancement. 
The \Hepsilon\ line is in emission for these PEBs but much less pronounced as the two EB examples in the sunspot moat. There is a hint of a weak absorption feature at \Hepsilon\ line center but it is not resolved. 

Figure~\ref{fig:timeseries} shows the temporal evolution of a QSEB in \Hbeta\ and \Hepsilon. 
The total lifetime of the QSEB is about 8~min as can be seen from the associated movie. 
The spectral profiles in Fig.~\ref{fig:timeseries} are taken from a fixed pixel location and the \Hbeta\ wing intensity increases by more than a factor 2 as the QSEB moves through this pixel location. 
The \Hepsilon\ line evolves from a weak emission feature in the early profiles (green and purple) to more enhanced for the pink and red profiles and finally absent in the last blue profile. 
The \Hepsilon\ line is stronger for the pink and red profiles but also the \CaH\ wing is more enhanced at these times. 
In the panels that show the difference between \Hepsilon\ and \CaH\ wing, the QSEB emission patch is clearly visible. 
In these panels we can also see that the strongest \Hepsilon\ emission is not right at the pixel location where \Hbeta\ wing gets strongest. The strongest \Hepsilon\ emission is sightly offset towards the upper left in the limb-ward direction. 
In the WB 4846~\AA\ images, no trace of the QSEB can be discerned which confirms the EB characteristic of being invisible in continuum radiation.


\begin{table*}
\caption{Statistical properties of QSEBs}
\label{table:stats}      
\centering                                     
\begin{tabular}{c c c c c c c }          
\hline\hline                      
                   & \multicolumn{2}{c}{QSEBs in \Hbeta\,(\citetalias{2022A&A...664A..72J})} & \multicolumn{2}{c}{QSEBs in \Hbeta}  & \multicolumn{2}{c}{QSEBs in \Hepsilon} \\
                                              & mean         &  median       & mean          & median       & mean         & median\\ 
\hline                   
Max. area [Mm$^2$] (pixels)\tablefootmark{a}  & 0.0277 (36)  &  0.0203 (26)  &  0.0270 (36)  &  0.0222 (30) & 0.0433 (58)  &  0.0334 (45)\\     
Lifetime [min] (frames)\tablefootmark{a}      & 1.65 (11)    &  1.14 (8)     &  1.80 (6)     &  1.20 (4)    & 1.62 (5)     &  1.20 (4)\\    
Max. brightness                               & 1.28         &  1.22         &  1.17         &  1.14        & 1.35         &  1.31\\     
\hline
\end{tabular}
\tablefoot{
\tablefoottext{a}{The values in parentheses are the nearest integer numbers.}
}
\end{table*}

\subsection{Quiet Sun Ellerman Bomb statistics}

The automated detection method we applied to the 16-Aug-2020 time series allowed us to measure various statistical properties of the QSEBs detected in the two lines. 
In total, over the full duration of the 24~min time sequence, we detected 961 QSEBs in the \Hbeta\ line and 1674 QSEBs in \Hepsilon. 
Of these, we found that 561 QSEBs were detected in both lines. 
Figure~\ref{fig:stats} presents distributions of the measured maximum area, lifetime, and maximum brightness of QSEBs detected in the two lines. 
Area and brightness are the maximum values over the lifetime of the QSEBs. 
The format of the figure allows for a direct comparison with the \Hbeta\ QSEB statistics in \citetalias{2022A&A...664A..72J} 
(their Fig.~6) and we conclude that our distributions are very similar as for their data. 
The mean and median values of the distributions are given in Table~\ref{table:stats}. The table also includes values from \citetalias{2022A&A...664A..72J}. 

Comparing the distributions between the two lines, we conclude that they are generally very similar. 
All distributions are positively skewed; the distributions have more weight toward the lower values and a tail toward the higher values. 
The distribution for maximum area in \Hepsilon\ is less skewed towards low values than in \Hbeta\ meaning that we find more larger area QSEBs in \Hepsilon\ (the median maximum area in \Hepsilon\ is 1.5 times larger).
The lifetime distributions are very similar with mean and median values that are almost identical. 
The values for maximum brightness are more difficult to compare between the two lines since they are measured differently. The QSEB brightness in \Hepsilon\ is measured against the \CaH\ blue wing, while in \Hbeta\ it is measured against the average far wing intensity in the local vicinity (like in \citetalias{2022A&A...664A..72J}). 
The maximum brightness in \Hepsilon\ has a sharp cutoff at a minimum value of 1.25. 
This cutoff was needed since we found that below this value, we obtained many clear false detections. 
We attribute these false detections due to neighboring spectral line blends that can suggest a weak emission feature in \Hepsilon. These blends are weak spectral lines from neutral species: \ion{Fe}{i}, \ion{Cr}{i}, and \ion{Ni}{i}.

Figure~\ref{fig:stats} includes joint probability distribution functions (JPDFs) with scatter plots between the three parameters. 
Like was presented in \citetalias{2022A&A...664A..72J}, 
both spectral lines show a general trend that QSEBs with a larger maximum area have a longer lifetime and are brighter as well. However, the scatter between these parameters is very large.
A comparison of the JPDFs that include the maximum area, panels (d) vs. (j) and panels (f) vs. (l), clearly shows that more QSEBs in \Hepsilon\ are larger. 

A subset of the QSEBs were detected in both spectral lines (561 QSEBs). 
Figure~\ref{fig:beta_vs_eps} presents statistics for the time difference when the QSEB first appeared in the two lines, their distance measured for their centers of gravity, and orientation. 
The measurement method was similar as for QSEBs in 
\citetalias{2022A&A...664A..72J} 
that showed both \Hbeta\ line core and line wing brightening, see their Fig.~5 for an illustration of the measurement. 
Most of the QSEBs appear first in \Hbeta\ ($\Delta t > 0$) while 21\%\ appear first in \Hepsilon.
All QSEBs have a spatial offset between the two lines, 8.5\%\ of the QSEBs have a distance $d$ smaller than 200~km. 
The direction of the offset is mostly in the limb-ward direction, we find that 70\%\ of the QSEBs in \Hepsilon\ are within $\pm$45\degr\ from their counterpart in \Hbeta. 
This means that the majority of the \Hepsilon\ QSEBs are closer to the limb. 
Figure~\ref{fig:beta_vs_eps}c presents the average propagation speed as measured from the distance $d$ and time difference $\Delta t$. 
The propagation speeds are comparable to what is found between \Hbeta\ line core and wing in 
\citetalias{2022A&A...664A..72J}: 
most speeds are $<$10~\kms.

As a test to see whether the QSEBs that are detected in both spectral lines are different as compared to the QSEBs that are detected only in \Hbeta, Fig.~\ref{fig:stat_comp} presents the \Hbeta\ statistical distributions separated for the two populations. 
The distributions for maximum area and brightness are very similar. For the distributions of lifetimes, it seems that there are more longer lived QSEBs that are detected in both lines. 
This comparison shows that QSEBs that are detected in both lines are not fundamentally very different from the QSEBs that are detected only in the \Hbeta\ line. 

Figure~\ref{fig:roi_offset} shows four different examples of QSEBs: two that are detected in both \Hbeta\ and \Hepsilon\, and two that are detected in only one spectral line. 
The QSEB in panels (a) appears first in \Hbeta\ and later in \Hepsilon. The QSEB in \Hepsilon\ is clearly offset in the limbward direction. The temporal evolution suggests an upward propagation speed of about 7~\kms. 
In example (b), the QSEB appears first in \Hepsilon\ and later in \Hbeta. There is no clear propagation speed but the QSEB in \Hepsilon\ is clearly offset in the limbward direction. 
The QSEB in (c) is clearly observed in \Hbeta\ but indiscernible in \Hepsilon\ while the QSEB in (d) is absent in \Hbeta\ but clearly present in \Hepsilon.
In the space-time diagram of example (d) there is an apparent upward propagation in the \Hepsilon\ line. 

Figure~\ref{fig:spatial_dist} presents the spatial distribution of QSEB detections over the full 24~min time series. 
QSEBs can be found throughout the FOV with more denser concentrations in the areas with stronger magnetic field. 
The spatial distribution of the photospheric magnetic field can be seen in the top panel of Fig.~\ref{fig:spatial_dist} that shows a map of the extreme values of $B_\mathrm{LOS}$. 
Since there are more than 1.7 times more QSEBs detected in \Hepsilon\ and more than half of the \Hbeta\ QSEBs are also detected in \Hepsilon, the QSEB distribution map is dominated by the blue dots of the \Hepsilon\ detections. 
The internetwork is also covered by QSEBs but the map shows small voids of approximately 3--6~Mm width. 
This means that there are small areas in the FOV in which no QSEBs appear during the 24~min duration of the time series. These empty areas are regions with weak magnetic fields with generally $|B_\mathrm{LOS}| < 50$~G.

Like in \citetalias{2022A&A...664A..72J}, 
we mark regions where over the duration of the time series, both polarities of significant strength ($|B_\mathrm{LOS}| > 50$~G) are present (blue contours). 
It illustrates that the occurrence of opposite polarities in close vicinity is very common, and we find QSEBs in and near these regions. 

\begin{figure}
    \includegraphics[width=0.48\textwidth]{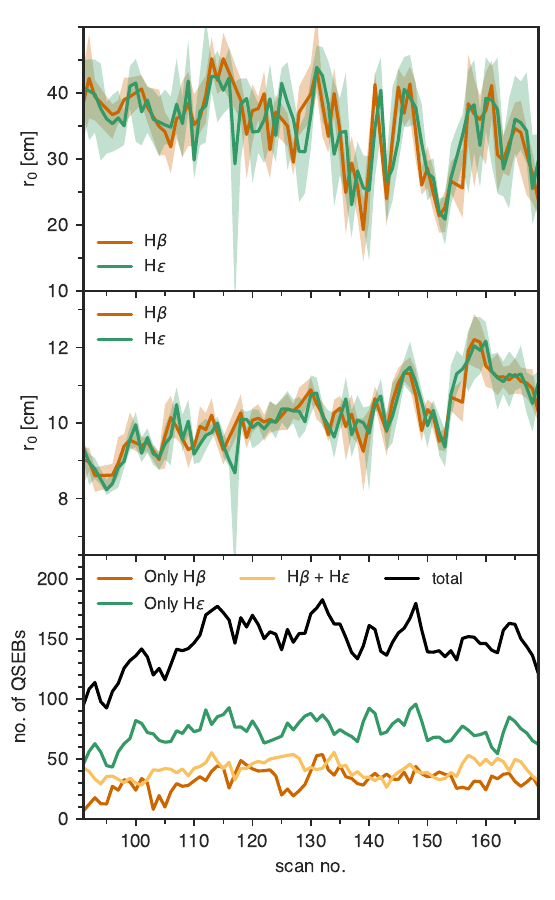}
    \caption{\label{fig:seeing}%
    The impact of seeing quality on the number of QSEB detections. The top two panels show the Fried's parameter $r_0$ as function of time (indicated as scan number in the 24 min time series, which started at 90). The top panel shows $r_0$ values that are a measure of the ground-layer seeing only. The middle panel shows $r_0$ values that measure the seeing over the whole atmosphere. The \Hbeta\ and \Hepsilon\ lines were recorded sequentially so there are two separate curves for the two spectral lines. The solid curves show the average $r_0$ values during the scan and the shaded area shows the range of $r_0$ values during the spectral scan. The bottom panel shows the number of QSEB detections per scan differentiated by: QSEBs detected only in \Hbeta, only in \Hepsilon, or both in \Hbeta\ and \Hepsilon. The black curve shows the total number of QSEB detections as the sum of these three curves. 
}
\end{figure}

Figure~\ref{fig:seeing} presents the number of QSEB detections in relation to the seeing quality. 
Like was found in \citetalias{2022A&A...664A..72J}, 
the number of QSEB detections is clearly correlated with the seeing quality: during the best seeing, the highest number of QSEBs are detected. 
The maximum number of QSEB detections is 182. This happened during a period of excellent seeing when the $r_0$ was peaking up to 50~cm. 
The highest number of \Hbeta\ QSEB detections is 94 (sum of \Hbeta\ only detections and both \Hbeta\ and \Hepsilon\ detections). The highest number of \Hepsilon\ detections is 140.

\section{Discussion}
\label{sec:discussion}

We used high quality observations to analyze EBs in the \Hepsilon\ line and compare with co-temporal \Hbeta\ and \Halpha\ observations. 
We conclude that the \Hepsilon\ line is well suited to study the EB phenomenon: it shows similar morphology and dynamics as in the traditional Balmer lines with the advantage of higher spatial resolution due to its shorter wavelength. 
This provides the potential to probe magnetic reconnection at very small spatial scale. 

In a ``textbook'' EB in an active region that occurred as the result of the coalescence of opposite magnetic polarity patches after strong magnetic flux emergence, we found similar flame-like morphology and rapid variability in the \Hepsilon\ line as in \Hbeta. 
In a small active region close to the limb, we see a tall and thin EB flame in both \Hepsilon\ and \Hbeta. 
The \Hepsilon\ line is clearly enhanced and for the observations where we have extended \CaH\ blue wing coverage, we see that the \Hepsilon\ line is clearly elevated as compared to the \CaH\ wing. 
In the penumbra of a sunspot, we find that the PEB emission in \Hepsilon\ is not as high as for EBs outside the sunspot. Still, the \Hepsilon\ line is a clear emission feature. 
Identification of weak \Hepsilon\ emission lines can be done effectively by subtraction with the opposite \CaH\ blue wing. 
For our quiet Sun observations this turned out to be an effective method to identify QSEBs. 

In a 24~min time series, we detected 1674 QSEBs in \Hepsilon\ and 961 QSEBs in \Hbeta. A subset of these, 561 QSEBs, were detected in both spectral lines (almost 60\%\ of the \Hbeta\ QSEBs). 
We found that the QSEB characteristics measured in \Hepsilon\ are not fundamentally different from the measurements in \Hbeta. The lifetime measurements are very similar. The QSEBs in \Hepsilon\ appear to be somewhat larger: the median of the maximum area distribution is 1.5 times larger in \Hepsilon. 
The spatial distribution of QSEBs in \Hepsilon\ is similar as in \Hbeta. In both lines, QSEBs are detected throughout the FOV with more denser concentrations in the network areas. Similar to our findings in 
\citetalias{2022A&A...664A..72J},  
we see small voids of width 3--6 Mm, where no QSEBs were detected. These voids are evenly distributed over the FOV, are generally regions with weak magnetic field and remind of the mesogranular pattern.
The majority of QSEBs are found in the vicinity of magnetic field concentrations. 
Most regions that have both magnetic polarities in close proximity also have QSEBs.
This supports the interpretation that QSEBs are effective markers of magnetic reconnection. 

We detected 961 \Hbeta\ QSEBs in the 24~min time series and
in \citetalias{2022A&A...664A..72J} we 
found 2809 QSEBs in a 60~min time series from 2019.
If we make a simple correction for the shorter time series, we see that we found 15\%\ fewer QSEBs. 
There are a few differences between the two data sets that could contribute to a lower detection number: in this 2020 data set, the two-line program resulted in a temporal cadence that is slightly double that of the 2019 data. 
The faster cadence 2019 data allowed for detecting more short lived QSEBs. 
Further, there are differences in the seeing quality. 
In \citetalias{2022A&A...664A..72J}, we
found a clear positive correlation between the seeing quality and the number of QSEB detections
and we find that also here (see Fig.~\ref{fig:seeing}). 
Both data sets have high image quality but the 2019 $r_0$ values for the full atmosphere are higher than the 2020 values for 14\%\ of the time steps (i.e., 59 time steps have $r_0 > 13$~cm). 
With more time steps having excellent seeing, more QSEBs could be detected in the 2019 data. 
%
Another difference is the observing angle: the 2020 data is 192\arcsec\ further away from the disk center and there is a 17\degr\ difference in observing angle. 
The solar photosphere is not a plane surface and the $\tau=1$ surface can be described as a corrugated landscape with granular ``hills'' and intergranular ``valleys''. One possible effect of different observing angles is that some small QSEBs that reside in the intergranular lanes could be hidden behind granules in the foreground in the 2020 data. 
Finally there could be a difference in number of QSEBs due to intrinsic differences between the two regions. 

We detected significantly more QSEBs in \Hepsilon\ than in \Hbeta\ (1.7 times more in \Hepsilon). 
During the best seeing moments, we detected up to 182 QSEBs (i.e., the sum of QSEBs detected only in \Hepsilon, only in \Hbeta, and detected both in \Hepsilon\ and \Hbeta). 
That is about a factor 1.5 times more QSEB detections than in 
\citetalias{2020A&A...641L...5J} 
and
\citetalias{2022A&A...664A..72J}. 
There, from a rough extrapolation of the highest detection number and the observation area, it was estimated that at any time, as many as 500,000 QSEBs might be present on the solar surface. 
The higher number of QSEB detections in \Hepsilon\ we find here suggest that that estimate can be increased to 750,000 QSEBs. 
We consider this a conservative increase of the estimate: for reasons discussed in the previous paragraph, we think it is likely that considerable larger number of QSEBs can be detected with a more dedicated \Hepsilon\ observation program with faster temporal cadence and with better seeing conditions. 

The formation of the traditional Balmer series lines, like \Halpha\ and \Hbeta, differs significantly from the formation of \Hepsilon\ in the solar atmosphere
\citep{2023A&A...677A..52K}. 
\Hepsilon\ is formed relative to the strong \CaH\ wing whereas \Halpha\ and \Hbeta\ are formed relative to the solar continuum.
The \CaH\ wing intensity at the \Hepsilon\ wavelength 
shows the reversed granulation intensity pattern that is formed higher
in the atmosphere than the solar optical continuum radiation. 
For \Hepsilon, the deepest we can view into the solar atmosphere is the reversed granulation layer. 
This is a height of about 300~km above the surface in the simulation that \citet{2023A&A...677A..52K}. 
analyzed. 
For \Halpha\ and \Hbeta, the deepest we can view is the solar photosphere represented by granulation observed in the line wings.
This could explain the limb-ward spatial offset between EB observations in \Hbeta\ and \Hepsilon, as \Hepsilon\ does not observe the EB structure below the reversed granulation layer.
This may further serve as one of the explanations of why some QSEBs are detected only in \Hbeta\ and some only in \Hepsilon: the QSEBs that are happening deepest in the atmosphere, may not be observable in \Hepsilon. QSEBs that are happening higher in the atmosphere, may not be observable in the \Hbeta\ wing while still present in \Hepsilon\ core. 
The time difference of many QSEBs that are detected first in \Hbeta\ and later in \Hepsilon\ suggests that often the magnetic reconnection starts first deep in the atmosphere and propagates upward in the atmosphere. Our measurements indicates that the typical propagation speed is below 10~\kms. 
We find that about a fifth of the QSEBs occur first in \Hepsilon. That suggests that there are also events for which the reconnection starts higher in the atmosphere and propagates downward. 
These findings are consistent with the time differences and propagation speeds measured between \Hbeta\ wing and core in 
\citetalias{2022A&A...664A..72J}. 
We find that on average, the maximum area of QSEBs detected in \Hepsilon\ is larger than in \Hbeta. 
Possibly, this is related to the higher formation height in \Hepsilon\ and we see the effect of magnetic structures fanning out to larger area with increasing height. 

There are other factors that could contribute to some QSEBs being detected in only one of the spectral lines. 
QSEBs evolve on very short time scales and some may have already decayed below the detection limit by the time the CHROMIS instrument was tuned to the other spectral line. 
Seeing variations are another factor that may be responsible for some of the non-detections. 
Furthermore, it may be that some weak emission in the other spectral line is in fact present but does not pass the thresholds of our detection method. 
In fact, we have done manual inspection of a number of single line QSEB detections and find weak QSEB features in the other spectral line. 

Most of the presented \Hepsilon\ EB spectral profiles show a weak absorption dip on top of the larger emission feature.
These weak absorption features imply that a part of the observed intensity comes from atmospheric layers above the EB reconnection site.
We observe the same morphological EB structures in the line core as in the wing of \Hepsilon.
That means that the layers above EBs are not optically thick enough to fully block the radiation coming from the EB and the absorption dip is formed under optically thin conditions.
That is different for the \Halpha\ and \Hbeta\ lines: both lines show chromospheric structures in the line core and the chromospheric fibrils are optically thick to \Halpha\ and (to somewhat lesser extent) \Hbeta\ line core radiation from the EB site below.
The fact that the chromosphere is optically thin in \Hepsilon\ can be seen in the \Hepsilon\ line core images in Fig.~\ref{fig:sunspot} and in Fig.~1 in 
\citet{2023A&A...677A..52K}: 
these images show a scene that is dominated by the \CaH\ wing and shows mostly inverse granulation. Only under larger magnification, some signature of thin chromospheric fibrils can be discerned in certain areas. These are the darkest fibrils in cotemporal \Halpha\ or \Hbeta\ line core images. 
This is also illustrated by the animation of the spectral line scan in Fig.~\ref{fig:textbook}.

We have demonstrated that the \Hepsilon\ line is well suited as a diagnostic of EBs. 
With its short wavelength, it arguably allows for the highest spatial resolution observations of magnetic reconnection in the solar atmosphere. 
This is of particular interest as a science case for the 4-m DKIST telescope 
\citep{2020SoPh..295..172R}, 
the upcoming EST telescope
\citep{2022A&A...666A..21Q}, 
and the third launch of the balloon-borne Sunrise observatory
\citep{2010ApJ...723L.127S} 
which has with its SUSI spectropolarimeter
\citep{2020SPIE11447E..AKF} 
access to multiple short-wavelength Balmer lines between 300 and 410~nm.

%
\begin{acknowledgements}
The Swedish 1-m Solar Telescope is operated on the island of La Palma
by the Institute for Solar Physics of Stockholm University in the
Spanish Observatorio del Roque de los Muchachos of the Instituto de
Astrof{\'\i}sica de Canarias.
The Institute for Solar Physics is supported by a grant for research infrastructures of national importance from the Swedish Research Council (registration number 2017-00625).
This research is supported by the Research Council of Norway, project numbers 250810, 
325491, 
and through its Centres of Excellence scheme, project number 262622.
K.K. acknowledges funding support by the European Research Council under ERC Synergy grant agreement No. 810218 (Whole Sun).
J.J. is grateful for travel support under the International Rosseland Visitor Programme. 
We are most grateful to Pit S{\"u}tterlin for his outstanding work to acquire the SST observations in service mode when international travel was impossible due to COVID19 restrictions and it was impossible for us to come to the telescope. 
We made much use of NASA's Astrophysics Data System Bibliographic Services.
\end{acknowledgements}


\begin{appendix}

\section{$k$-means clustering and QSEB detections}
\label{app:kmeans}

\begin{figure*}
    \includegraphics[width=\textwidth]{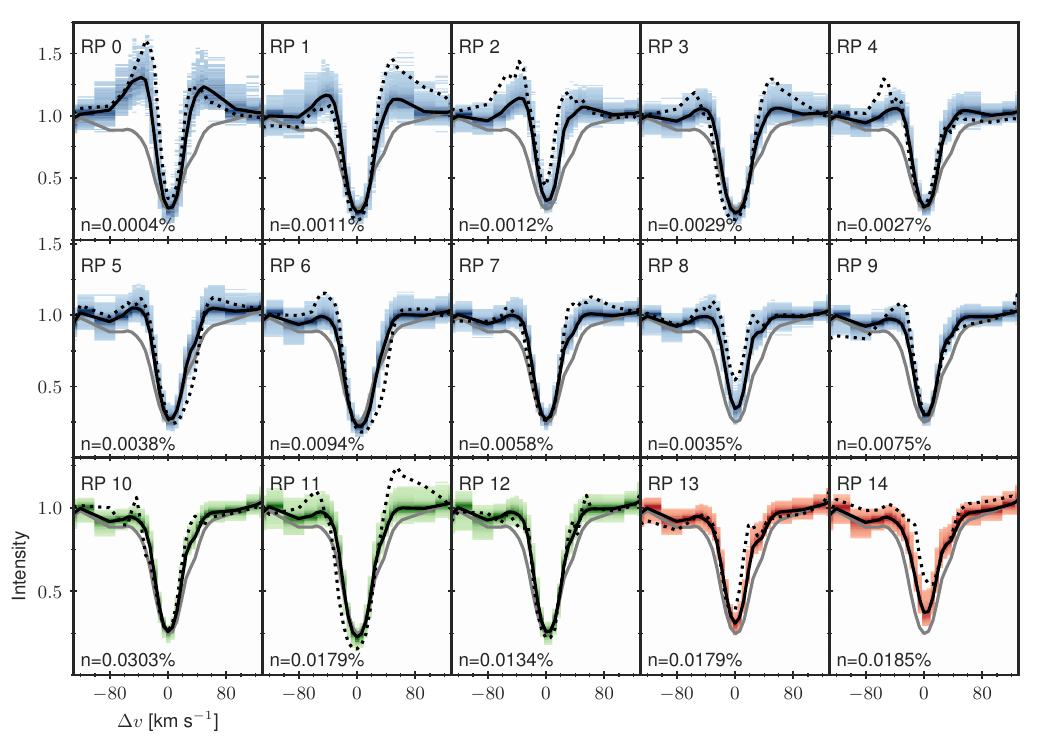}
    \caption{\label{fig:RPs_hbeta} 
Fifteen RPs from the $k$-means clustering of the \Hbeta\ line that are identified as signature of QSEB. The black lines show RPs whereas shaded colored areas represent density distribution of \Hbeta\ spectra within a cluster; darker shades indicate higher density. Within a particular cluster, the \Hbeta\ profile that is farthest (measured in euclidean distance) from the corresponding RPs is shown by the black dotted line. As reference, the average quiet Sun profile (gray line) is plotted in each panel. RPs 0--9 (blue) show the typical EB-like \Hbeta\ profiles, i.e., significantly enhanced wings and unaffected line core, while RPs 10--12 (green) display weak enhancement in the wings. RPs 13 and 14 (red) show intensity enhancement in the line core. The differentiation into three groups of QSEB clusters is similar as in \citetalias{2022A&A...664A..72J}. 
The parameter $n$ represents the number of spectral profiles in a cluster as percentage 
of the total of $\sim 1.7\times10^{9}$ spectra.}
\end{figure*}

\begin{figure*}
    \includegraphics[width=\textwidth]{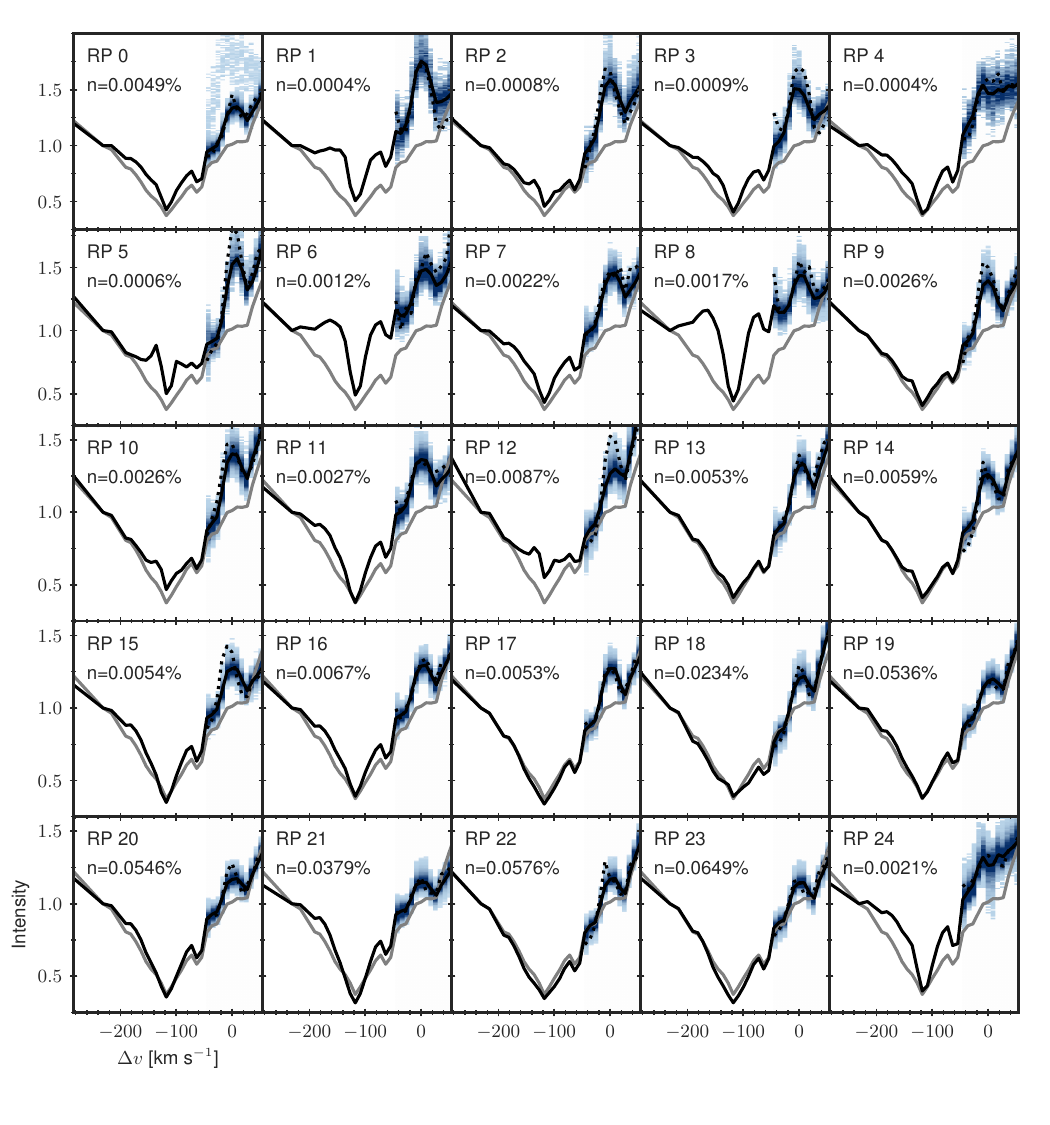}
    \caption{\label{fig:RPs_heps} 
    Twenty five RPs (solid black lines) from the $k$-means clustering of the \Hepsilon\ line that are identified as signature of QSEB.  The $k$-means clustering has been done on 13 line positions around \Hepsilon\ nominal line core. The shaded colored areas represent density distribution of \Hepsilon\ spectra within a cluster. The black dotted line shows the \Hepsilon\ profile that is farthest (measured in euclidean distance) from the corresponding RP within a particular cluster. The average quiet Sun profile (gray line) is plotted in each panel as reference. The parameter $n$ represents the number of spectral profiles in a cluster as percentage of the total of $\sim 1.7\times10^{9}$ spectra.
}
\end{figure*}

Figures~\ref{fig:RPs_hbeta} and \ref{fig:RPs_heps} show the representative profiles (RPs) from the $k$-means clustering of the \Hbeta\ and \Hepsilon\ lines that are identified as signature of QSEB.
The clustering was done for 81 time steps of the 16-Aug-2020 quiet Sun data having 1686 $\times$ 1000 pixels per time step. 
We used 100 clusters for both the clustering of the \Hbeta\ and \Hepsilon\ spectra. 
The \Hbeta\ QSEB clusters in Fig.~\ref{fig:RPs_hbeta} are differentiated into three groups following 
\citetalias{2022A&A...664A..72J}. 
The blue clusters (RPs 0--9) are the characteristic EB profiles with significantly enhanced wings and largely unaffected line core.
Each \Hbeta\ QSEB event is required to have profiles from these clusters. 
Pixels that have profiles from the green clusters (RPs 10-12, weakly enhanced wings) or from the red clusters (RPs 13 and 14, enhanced line core) are added to the event if they are connected in space and time to the pixels with blue cluster profiles. 
More details on the detection of \Hbeta\ QSEBs are provided in 
\citetalias{2022A&A...664A..72J}. 

\begin{figure}
    \includegraphics[width=0.5\textwidth]{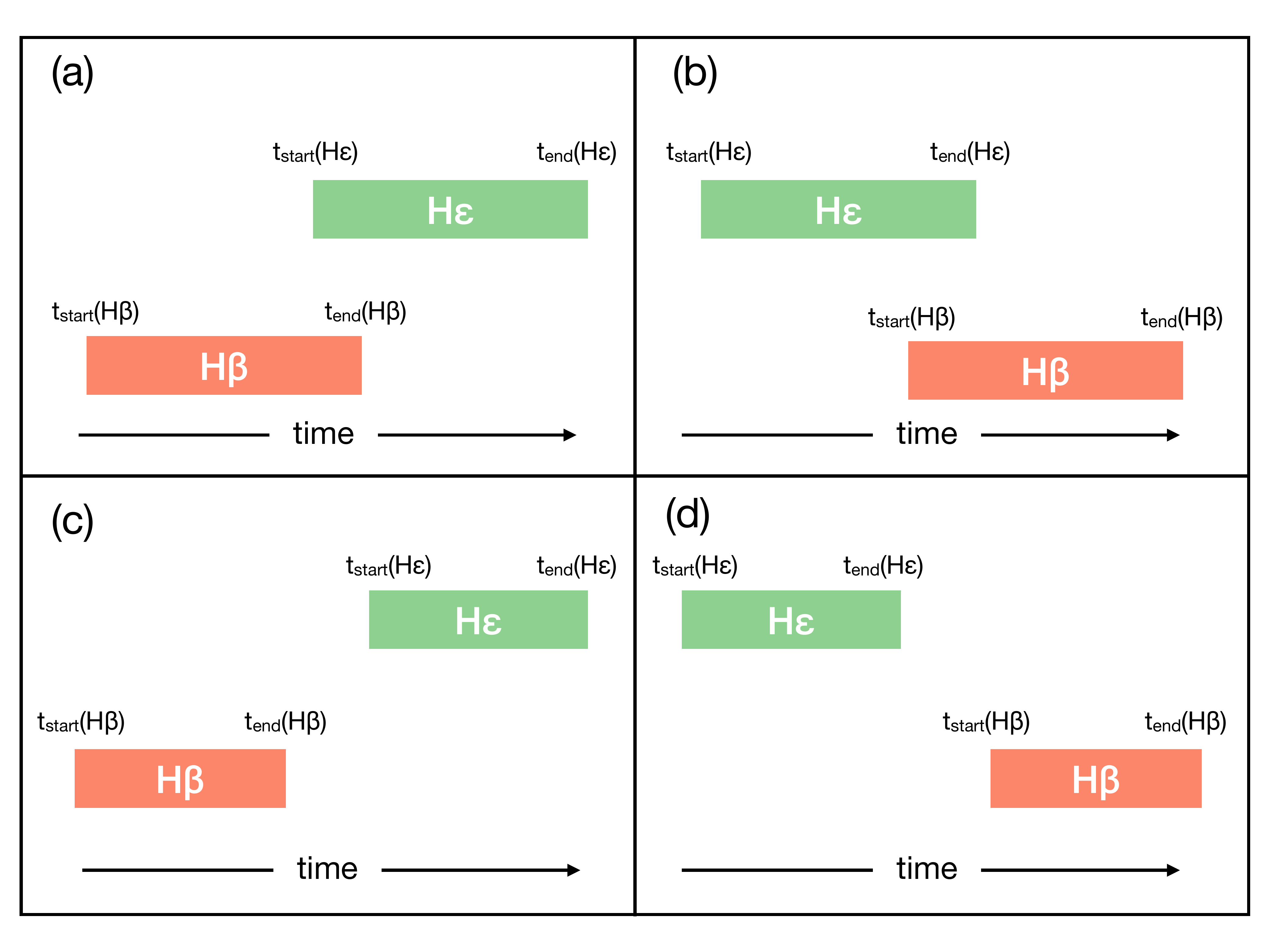}
    \caption{\label{fig:hbeta_heps_conn} %
    Illustration of connecting \Hbeta\ and \Hepsilon\ QSEB detections in time. Four different cases are shown with the duration a QSEB detection continues illustrated with colored bars. The spatial connection is not illustrated but it is assumed that the \Hbeta\ and \Hepsilon\ detections are spatially connected. The cases in panel (a) and (b) have temporal overlap and are therefore clearly temporally connected. The cases in panels (c) and (d) have a time gap between detections and need to satisfy a temporal condition in order to be considered as a connected event.  
    }
\end{figure}

Figure~\ref{fig:hbeta_heps_conn} illustrates the procedure for connecting \Hbeta\ and \Hepsilon\ QSEB detections in time. 
For all four cases, it is assumed that the \Hbeta\ and \Hepsilon\ detections are connected in space: the centroids of all pairs are separated by less than 500~km. 
In panel (a), the \Hbeta\ QSEB detection starts before the \Hepsilon\ detection but the \Hepsilon\ detection starts before the \Hbeta\ detection ends ($t_\mathrm{start}(\Hepsilon) < t_\mathrm{end}(\Hbeta)$).
The QSEB is detected for at least one time step in both \Hbeta\ and \Hepsilon\ so this is a clear connected QSEB. 
The case in panel (b) is also a clear connected QSEB since the later \Hbeta\ QSEB detection starts before the end of the \Hepsilon\ detection. 
For the cases in panels (c) and (d), there is a time gap between the QSEB detections: the \Hbeta\ and \Hepsilon\ detections are close in time but do not share any common time step.
The maximum allowed time gap is 9 time steps or 162 s (i.e., for (c): $t_\mathrm{start}(\Hepsilon) - t_\mathrm{end}(\Hbeta) \le$ 162~s and for (d): $t_\mathrm{start}(\Hbeta) - t_\mathrm{end}(\Hepsilon) \le$ 162~s).

The 500~km threshold for spatially connecting was found to be a reasonable limit to avoid many ambiguous connections. 
By varying the spatial distance threshold we found that for larger distances above 500~km, the number of connected events increased rapidly and the statistical distributions for orientation and time difference in Fig.~\ref{fig:beta_vs_eps} changed significantly. 

\end{appendix}

\end{document}